\documentstyle[12pt]{article}

\begin{document}

\newcommand{\lesssim}{\mbox{\tiny$\mbox{\normalsize$<$}\atop
\mbox{\normalsize$\sim$}$}}

\title{On the Selfconsistent Theory of Josephson Effect in Ballistic
Superconducting Microconstrictions}
\author{Malek Zareyan~${}^{a}$ Yu.A.Kolesnichenko~${}^{b}$ and A.N.\ Omelyanchouk~$%
{}^{b}$ \\
{\small {\em ${}^{a}$ Institute for Advanced Studies in Basic Sciences,}} \\
{\small {\em 45195-159, Gava Zang, Zanjan, Iran}} \\
{\small {\em ${}^{b}$ B.Verkin Institute for Low Temperature Physics and
Engineering,}} \\
{\small {\em National Academy of Sciences of Ukraine,}} \\
{\small {\em 47 Lenin Ave., 310164 Kharkov, Ukraine}}}
\date{}
\maketitle

\begin{abstract}
The microscopic theory of current carrying states in the ballistic
superconducting microchannel is presented. The effects of the contact length 
$L$ on the Josephson current are investigated. For the temperatures $T$
close to the critical temperature $T_{c}$ the problem is treated
selfconsistently, with taking into account the distribution of the order
parameter $\Delta \left( {\bf r}\right) $ inside the contact. The closed
integral equation for $\Delta $ in strongly inhomogeneous microcontact
geometry ($L\lesssim \xi _{0}$, $\xi _{0}$ is the coherence length at $T=0$
) replaces the differential Ginzburg-Landau equation. The critical current $%
I_{c}(L)$ is expressed in terms of solution of this integral equation. The
limiting cases of $L\ll \xi _{0}$ and $L\gg \xi _{0}$ are considered. With
increasing length $L$ the critical current decreases, although the ballistic
Sharvin resistance of the contact remains the same as at $L=0.$ For ultra
short channels with $L\lesssim a_{D}$ ($a_{D}\sim v_{F}/\omega _{D}$ , $%
\omega _{D}$ is the Debye frequency) the corrections to the value of
critical current $I_{c}(L=0)$ are sensitive to the strong coupling effects.
\end{abstract}

\section{Introduction}

Weak superconducting links \cite{Jo} include the tunnel structures $S-I-S$
(superconductor-insulator-superconductor) and the contacts with direct
conductivity, $S-N-S$ ($N$ is the normal layer) and $S-c-S\,$ ($c$ is a
geometrical constriction) . Superconducting constrictions can be modelled as
the orifice with diameter $d$ in a inpenetrable sheet for electrons between
two superconducting half spaces (point contact), or as a narrow channel with
length $L\,$ in a contact with superconducting banks (microbridge).
Aslamazov and Larkin \cite{AL} have shown on the basis of a solution of the
Ginzburg-Landau (GL) equations that in the dirty limit and for small
constriction's sizes $L,d\ll \xi (T)$ ( $\xi (T)$ is GL coherence length)
the $S-C-S\,$ contact can be described by a Josephson model with the
current-phase relation

\begin{equation}
I=I_{c}\sin {\varphi }, I_{c}=\pi \Delta _{0}^{2}(T)/(4eR_{N}T_{c}),
\label{IAL}
\end{equation}
where $I_{c}$ is the Josephson critical current, $\Delta _{0}$ the absolute
value of the order parameter in the bulk banks, $T_{c}$ the critical
temperature and $R_{N}$ the normal-state resistance of dirty microbridge.
The critical current of microbridge (\ref{IAL}) depends on the bridge length
as $I_{c}\sim 1/L$. The expression (\ref{IAL}) is valid within the domain of
applicability of GL approach, {\it i.e.} for temperatures $T$ close to $%
T_{c} $ and $L,d\gg \xi _{0}$ ($\xi _{0}\simeq v_{F}/T_{c}$ is the coherence
length at $T=0,v_{F}$ is the Fermi velocity).

The present level of technology have made it possible to study the
ultrasmall Josephson weak links with the dimensions up to interatomic size.
For example, it can be nanosized microchannels produced by means of a
scanning tunneling microscope \cite{Poza} or point contacts and
microchannels obtained by using the mechanical controllable break technique 
\cite{Mul}, \cite{Na, Na2} . The microchannels between two superconductors
can also arise spontaneously as microshorts in tunnel junctions\cite{Ya},
with the length $L$ determined by the thickness of an insulator layer. The
value of the critical current $I_{c}$ of such microshorts presents the
special interest in the case of tunnel structures based on high-$T_{c}$
metalloxide compounds. The case of small microconstrictions with dimensions
of the order or smaller than coherence length $\xi _{0}$ , when the
expression (\ref{IAL}) for the critical current $I_{c}\sim 1/L$ is not
valid, requires the microscopic consideration even for $T$ near $T_{c}$.
Such microscopic theory of stationary Josephson effect in microconstrictions
was developed in Ref. \cite{KO} for the ballistic channel of zero length $L=0
$ in the model of the orifice of diameter $d<<\xi _{0}$. The Josephson
current in this case is given by:

\begin{equation}
I=\frac{\pi {\Delta }_{0}(T)}{eR_{0}}\sin {\varphi /2}\tanh {\frac{{\Delta }%
_{0}(T)\cos {\varphi /2}}{2T},}-\pi <\varphi <\pi ,  \label{IKO}
\end{equation}

\begin{equation}
R_{0}^{-1}=\frac{1}{2}Se^{2}v_{F}N(0),  \label{RSh}
\end{equation}
where $S=\pi d^{2}/4$ is the contact cross-section area, $N(0)=mp_{F}/(2{\pi 
}^{2})$ is electron density of states at the Fermi surface. At temperatures $%
T_{c}-T\ll T_{c}$ expression (\ref{IKO}) coincides with the Aslamazov-Larkin
result, Eq. (\ref{IAL}), in which instead of the normal resistance $R_{N}\,$
for dirty metal, the ballistic Sharvin resistance \cite{Sha}\thinspace $%
R_{0}\,$(\ref{RSh}) is substituted.

The purpose of this work is to present a microscopic theory of current
carrying states in the ballistic microbridges having the length $L$
arbitrary in the scale of the coherence length $\xi _{0}$. We have
investigated the dependence of Josephson critical current on the ratio $%
L/\xi _{0}$ and analyzed the transition from the case of $I_{c}(L=0)$ (\ref
{IKO}) to $I_{c}\sim 1/L$ (\ref{IAL}) with increasing of the length $L$.

In Sec.2 we formulate the model of a microbridge and the microscopic
equations for Green's functions with boundary conditions at the bridge
edges. When the effects on the critical current of the microconstriction's
length are studied, the crucial point, as always in space inhomogeneous
superconducting state, consists in the self-consistent treatment of the
order parameter distribution $\Delta ({\bf r})$ inside the weak link. In
Sec.3 the closed integral equation for the order parameter $\Delta $ in
microchannel is derived for temperatures near $T_{c}$, which in strongly
inhomogeneous ($L\sim $ $\xi _{0}$) microcontact geometry replaces the
differential GL equation. The critical current $I_{c}(L)$ is expressed in
terms of the solution of this integral equation. The limiting cases of $L\ll
\xi _{0}$ and $L\gg \xi _{0}$ are considered. We shall show that beside the
characteristic scale $\xi _{0}$ the length $a_{D}\simeq v_{F}/\omega _{D}$ ($%
\omega _{D}$ is Debye frequency) appears in the case of ultra small channel.
The length $L\sim a_{D}$ is the length at which the frequency of the
ballistic flight of electron from one bank to another becomes comparable
with the frequency $\omega _{D}$ , which characterizes the retardation of
the electron-phonon interaction. In conventional superconductors the value
of the coherence length $\xi _{0}$ is about $10^{-4}cm$ and is much lager
than $a_{D}\sim 100\AA $. But in high-$T_{c}$ metalloxide compounds we have
the situation with $\xi _{0}$ comparable with $a_{D}.$ Thus, in high-$T_{c}$
compounds the critical current of the contact with dimensions $\sim
a_{D}\sim \xi _{0}$ will be sensitive to the effects of strong coupling.

\section{Model and basic equations}

We consider the model of a contact in the form of a filament (narrow
channel) that joins two superconducting half-spaces (massive banks) (Fig.1).
The length $L$ and the diameter $d$ of the channel are assumed to be large
as compared with the Fermi wavelength $\lambda _{F}$, so we can apply the
quasi-classical approximation. In the ballistic case, we proceed from the
quasi-classical Eilenberger equation for the energy-integrated Green's
function \cite{Ei}:

\begin{equation}
{\bf v_{F}}.\frac{\partial }{\partial {\bf r}}\hat{G}+[\omega \hat{\tau}_{3}+%
\hat{\Delta},\hat{G}]=0,  \label{eq1}
\end{equation}
where

\[
\hat{G}(\omega ,{\bf v_F},{\bf r})=\left(
\begin{array}{cc}
g_\omega & f_\omega \\
f_\omega ^{+} & -g_\omega
\end{array}
\right)
\]
is the matrix Green function, which depends on the Matsubara frequency $%
\omega $, the electron velocity on the Fermi surface ${\bf v_F}$ and the
spatial variable ${\bf r}$;

\[
\hat{\Delta}({\bf r})=\left(
\begin{array}{cc}
0 & \Delta \\ 
\Delta ^{\ast } & 0
\end{array}
\right) 
\]
is the superconducting pair potential; $\hat{\tau}_{i}$ ($i=1,2,3$) are
Pauli matrices. Equation for matrix Green's function (\ref{eq1}) is
supplemented by the normalization condition \cite{La}

\begin{equation}
\hat{G}^{2}=1.  \label{G^2}
\end{equation}

The off-diagonal potential $\Delta ({\bf r)}$ must be determined from the
self-consistency equation:

\begin{equation}
\Delta ({\bf r})=\lambda 2\pi T\sum_{\omega >0}<f>  \label{delta0}
\end{equation}
in which $<...>$ stands for averaging over directions of ${\bf v_{F}}$ on
the Fermi surface and $\lambda $ is the electron-phonon coupling constant.
In the BCS model the summation over $\omega $ contains the cutoff on the
frequency $\omega _{D},$ which is of the order of the Debye frequency .

The equations (\ref{eq1}) and (\ref{delta0}) are supplemented by the values
of Green's functions and $\Delta $ in the bulk superconductors $S_{1}$ and $%
S_{2}$ far from the channel ends :

\begin{equation}
\hat{G}_{1,2}=\frac{\omega \hat{\tau}_{3}+\hat{\Delta}_{1,2}}{\Omega },\hat{%
\Delta}_{1,2}=\Delta _{0}(\cos (\varphi /2)\hat{\tau}_{1}\pm \sin (\varphi
/2)\hat{\tau}_{2}),  \label{bound0}
\end{equation}
thus the phase $\varphi $ is the total phase difference on the contact. Also
we have to determine the boundary conditions concerning the reflection of
the electrons from the surface of the superconductors ${\bf r}_{S}$. For
simplicity we will assume that at ${\bf r}_{S}$ electrons undergo the
specular reflection. Then for quasiclassical Green's function we have the
boundary condition (Ref.\cite{KO}):

\begin{equation}
G({\bf v_{F}},{\bf r}_{S})=G({\bf v_{F}}^{\prime },{\bf r}_{S})
\end{equation}
in which ${\bf v_{F}}$ and ${\bf v_{F}}^{\prime }$ are the velocities of the
incident and specular reflected electron. These velocities are related by
the conditions, which conserve the component of ${\bf v}_{F}$ parallel to
the reflecting surface ${\bf r}_{S}$ and changes the sign of the normal
component.

The solutions of the equations (\ref{eq1}) and (\ref{delta0}) allow us to
calculate the current density ${\bf j}$:

\begin{equation}
{\bf j}(r)=-4i\pi eN(0)T\sum_{\omega >0}<{\bf v_{F}}g_{\omega }>.  \label{I}
\end{equation}

In the case of the microconstriction shown in Fig.1, under the conditions $%
d\ll \xi _{0}$ and $L\gg d$ ($d$ is the contact diameter), inside the
filament we can solve the one-dimensional Eilenberger equations with $\Delta
=\Delta (z)$ $.$ The banks of the bridge are equivalent here to certain
boundary conditions for Green's function $\hat{G}(v_{z,}z)$ at points $z=\pm 
\frac{L}{2}$. Following the procedure, which was described in Refs. \cite{KO}%
, one can find the Green's functions at the end points ($z=\pm \frac{L}{2}$)
from the general solutions of Eq.(\ref{eq1}) in superconducting half-spaces $%
S_{1}$ and $S_{2}$ with conditions (5). They are given by

\begin{equation}
\begin{array}{c}
\hat{G}(z=\mp \frac{L}{2})=\hat{G}_{1,2}+A_{1,2}[\Delta _{0}\hat{\tau}%
_{3}-(\omega \cos (\varphi /2)+i\eta \Omega \sin (\varphi /2))\hat{\tau}%
_{1}\mp  \\
(\omega \sin (\varphi /2)-i\eta \Omega \cos (\varphi /2))\hat{\tau}_{2}],
\end{array}
\label{bound}
\end{equation}
where $\Omega =\sqrt{\omega ^{2}+{\Delta _{0}}^{2}}$, $\eta =sign(v_{z})$.
The arbitrary constants $A_{1,2}$ must be determined by matching of these
boundary conditions with the solution for $\hat{G}(v_{z,}z)$ inside the
channel.

Taking the off-diagonal components of Eq.(\ref{eq1}), we have the following
first-order differential equations for the anomalous Green's functions:

\begin{equation}
\begin{array}{c}
v_{z}\frac{df_{\omega }}{dz}+2\omega f_{\omega }=2\Delta (z)g_{\omega }, \\
-v_{z}\frac{df_{\omega }^{+}}{dz}+2\omega f_{\omega }^{+}=2\Delta ^{\ast
}(z)g_{\omega }.
\end{array}
\label{eq2}
\end{equation}
The normal Green's function $g_{\omega }$, as follows from condition (\ref
{G^2}), is expressed in terms of $f_{\omega }$ and $f_{\omega }^{+}$ :

\begin{equation}
g_{\omega }=\sqrt{1-f_{\omega }f_{\omega }^{+}}.  \label{g_om}
\end{equation}

From equations (\ref{I}), (\ref{delta0}),(\ref{eq2}) and \ref{g_om} one
obtains the symmetry relations

\begin{equation}
f_{\omega }^{+}(v_{z},z)=(f_{\omega }(-v_{z},z))^{\ast },\Delta ^{\ast
}(z)=\Delta (-z)  \label{f+}
\end{equation}
and the current conservation inside the channel $dj/dz=0$ .

\section{Josephson current and order parameter distribution in
superconducting microchannel}

In present paper we consider the case of temperatures $T$ close to the
critical temperature $T_{c}$. Near the phase transition curve the order
parameter $\Delta _{0}(T)$ in the banks is small. In order to find the
Josephson current in the lowest order on $\Delta _{0}$ we linearize the
equations (\ref{eq2} ) on $\Delta $ and obtain $f_{\omega }\sim \Delta
_{0}(T)$, $g_{\omega }\simeq 1-1/2f_{\omega }f_{\omega }^{+}\sim 1-O(\Delta
_{0}^{2})$, $j\sim \Delta _{0}^{2}$. The equation for $f_{\omega }$ near $%
T_{c}$ takes the form 
\begin{equation}
v_{z}\frac{df_{\omega }}{dz}+2\omega f_{\omega }=2\Delta (z),  \label{eq3}
\end{equation}
with linearized boundary conditions (\ref{bound} )

\begin{equation}
\begin{array}{c}
f_{\omega }(v_{z}>0,z=-L/2)=\frac{\Delta _{0}}{\omega }e^{-i\frac{\varphi }{2%
}}, \\ 
f_{\omega }(v_{z}<0,z=+L/2)=\frac{\Delta _{0}}{\omega }e^{+i\frac{\varphi }{2%
}}.
\end{array}
\label{bound1}
\end{equation}
Its solution for arbitrary function $\Delta (z)$ is given by

\begin{equation}
f_{\omega }(v_{z},z)=\frac{\Delta _{0}}{\omega }e^{-i\eta \frac{\varphi }{2}%
}e^{-\frac{2\omega }{v_{z}}(z+\eta L/2)}+e^{-\frac{2\omega }{v_{z}}%
z}\int\limits_{-\eta L/2}^{z}dz^{\prime }\frac{2\Delta (z^{\prime })}{v_{z}}%
e^{\frac{2\omega }{v_{z}}z^{\prime }}.  \label{fomega}
\end{equation}
The Green's function $f_{\omega }^{+}(v_{z},z)$ is obtained from expression (%
\ref{eq3}) with the help of relations (\ref{bound1} ).

Substituting function $f_{\omega }(v_{z},z)$ (\ref{fomega}) in the
self-consistency equation (\ref{delta0}), we obtain the integral equation
for the space-dependent order parameter inside the contact

\begin{equation}
\Delta (z)=A(z)+\int\limits_{-L/2}^{L/2}dz^{\prime }\Delta (z^{\prime
})K(\left| z-z^{\prime }\right| ),  \label{eqdelta}
\end{equation}
where

\begin{equation}
A(z)=\lambda 2\pi T\sum\limits_{\omega >0}\frac{\Delta _{0}}{\omega }%
\left\langle e^{-\frac{\omega L}{v_{z}}}\cosh (\frac{2\omega z}{v_{z}}+i%
\frac{\varphi }{2})\right\rangle _{v_{z}>0},  \label{A}
\end{equation}

\begin{equation}
K(z)=\lambda 2\pi T\sum\limits_{\omega >0}\left\langle \frac{1}{v_{z}}e^{-%
\frac{2\omega }{v_{z}}z}\right\rangle _{v_{z}>0}.  \label{K}
\end{equation}
The averaging $<...>_{v_{z}>0}$ denotes $\ <F(v_{z}=v_{F}\cos \theta
)>_{v_{z}>0}=\int_{0}^{1}d(\cos \theta )F(\cos \theta ).$

In the case of strongly inhomogeneous microcontact problem the integral
equation for the order parameter $\Delta $ replaces the differential
Ginzburg-Landau equation . It contains the needed boundary conditions at the
points of contact between the filament and the bulk superconductors. Some
general properties of the solution $\Delta (z)$ of Eq.(\ref{eqdelta} )
follow from the form of the functions ( \ref{A}) and (\ref{K}). Let us write 
$\Delta (z)$ in the form

\begin{equation}
\Delta (z)=\Delta _{0}(T)(\cos \frac{\varphi }{2}+iq(z)\sin \frac{\varphi }{2%
})  \label{delta}
\end{equation}
and substitute it in the Eq.( \ref{eqdelta}). For function $q(z)$ we obtain
the equation

\begin{equation}
q(z)=b(z)+\int\limits_{-L/2}^{L/2}dz^{\prime }q(z^{\prime })K(\left|
z-z^{\prime }\right| ),  \label{q(z)}
\end{equation}
with $K(z)$ defined by ( \ref{K}) and the new out-integral function $b(z)$

\begin{equation}
b(z)=\lambda 2\pi T\sum\limits_{\omega >0}\frac{1}{\omega }\left\langle e^{-%
\frac{\omega L}{v_{z}}}\sinh (\frac{2\omega z}{v_{z}})\right\rangle
_{v_{z}>0}.  \label{b}
\end{equation}
In obtaining the Eqs(\ref{q(z)}) we have used the relation

\begin{equation}
\lambda 2\pi T\sum\limits_{\omega >0}^{\omega_{D}}\frac{1}{\omega }=1,
\ \ {\rm for}\ \ T\rightarrow T_{c}.  \label{lambda}
\end{equation}

It follows from (\ref{q(z)}), (\ref{K}) and (\ref{b}) that function $q(z)$
has such properties:

{\it i) }function $q(z)$ is real,

{\it ii) }$q(z)$ does not depend on the phase $\varphi $,

{\it \ iii)} $q(-z)=-q(z),q(0)=0.$

Thus, the value of the order parameter $\Delta $ at the center of the
contact always equals to $\Delta _{0}(T)\cos \frac{\varphi }{2}$ . Also, the
universal phase dependence of $\Delta (z,\varphi )$, which is determined by (%
\ref{delta}) and {\it i)-iii), }leads (see below) to the sinusoidal
current-phase dependence $j=j_{c}\sin \varphi $. It is emphasized, that
these general properties of the ballistic microchannel (within the
considered case of ''rigid'' boundary conditions (\ref{bound}) and
temperatures close to $T_{c}$) {\it does not depend }on{\it \ }the contact
length $L$, in particular, on the ratio of $L/\xi _{0}$.

Now we are going to obtain the Josephson current in the system. To calculate
the total current $I=Sj$ flowing through the channel at given phase
difference $\varphi ,$ we use the equation for the current density (\ref{I})
and the obtained above anomalous Green's function $f_{\omega }$ (\ref{fomega}%
). The normal Green's function $g_{\omega }$ (\ref{g_om}) in the second
order on $\Delta _{0}(T)$ equals to $g_{\omega }(v_{z},z)=1-\frac{1}{2}%
f_{\omega }(v_{z},z)(f_{\omega }(-v_{z},z))^{\ast }$. It is convenient to
calculate the current density at the point $z=0$. By using the expression
for $\Delta (z)$ (\ref{delta}), we obtain the general formula for the
Josephson current $I(\varphi )$ in terms of function $q(z)$ $:$

\begin{equation}
I(\varphi )=I_{c}\sin \varphi ,  \label{I1}
\end{equation}

\begin{equation}
I_{c}=I_{0}\frac{16T^{2}}{v_{F}}\sum\limits_{\omega >0}\left[ \frac{1}{%
\omega ^{2}}\left\langle v_{z}e^{-\frac{\omega L}{v_{z}}}\right\rangle
_{v_{z}>0}+\frac{2}{\omega }\int\limits_{0}^{L/2}dzq(z)\left\langle e^{-%
\frac{2\omega }{v_{z}}z}\right\rangle _{v_{z}>0}\right] .  \label{Ic}
\end{equation}
Here $I_{0}=\pi \Delta _{0}^{2}(T)/(4eR_{0}T_{c})$ is the critical current
at $L=0$. It coincides with the result of Ref. \cite{KO} for the orifice (%
\ref{IKO}) at $T$ near $T_{c}$. Expression (\ref{Ic} ) jointly with equation
(\ref{q(z)} ) for $q(z)$ describes the dependence of the critical current on
the contact length $I_{c}(L)$. It is valid for arbitrary value of the ratio $%
L/\xi _{0}$. Note, that in considered here case $T\rightarrow T_{c}$, we
have the relation $\xi _{0},L\ll \xi (T).$

Let us introduce the dimensionless quantities

\begin{equation}
x=z/L,\ell =\frac{\pi T_{c}L}{v_{F}},\frac{\omega }{\pi T_{c}}=2n+1,J_{c}=%
\frac{I_{c}}{I_{0}}.  \label{dless}
\end{equation}
In reduced units (\ref{dless}), after taking the average $<...>_{v_{z}>0}$,
the equations for $q(x)$ and $J_{c}$ take the form

\begin{equation}
q(x)=b(x)+\ell \int\limits_{-1/2}^{1/2}dx^{\prime }q(x^{\prime })K(\left|
x-x^{\prime }\right| ),  \label{q(x)1}
\end{equation}

\begin{equation}
\begin{array}{c}
J_{c}=\frac{8}{\pi ^{2}}\sum\limits_{n=0}^{N}\{\frac{\exp [-\ell
(2n+1)][1-\ell (2n+1)]}{(2n+1)}-\ell ^{2}Ei[-\ell (2n+1)]+ \\ 
+4\ell \int\limits_{0}^{1/2}dxq(x)[\frac{\exp [-2\ell (2n+1)x]}{(2n+1)}%
+2\ell xEi[-2\ell (2n+1)x]\},
\end{array}
\label{Ic1}
\end{equation}
where

\begin{equation}
\begin{array}{c}
b(x)=\lambda \sum\limits_{n=0}^{N}\{\frac{2\exp [-\ell (2n+1)]\sinh [2\ell
(2n+1)x]}{(2n+1)}+\ell (2n+1)(1-2x)Ei[-\ell (2n+1)(1-2x)]+ \\ 
-\ell (2n+1)(1+2x)Ei[-\ell (2n+1)(1+2x)]\},
\end{array}
\label{b1}
\end{equation}

\begin{equation}
K(x)=-2\lambda \sum\limits_{n=0}^{N}Ei[-2\ell (2n+1)x].  \label{K1}
\end{equation}
Function $Ei(x)=\int_{-\infty }^{x}\frac{\exp (t)}{t}dt$ is the integral
exponent. The upper limit $N$ in the sums over $n$ is related to the cutoff
frequency $\omega _{D}$ in the BCS model, $N\simeq \omega _{D}/T_{c}$. The
value of coupling constant $\lambda $ is related to $N$ by Eq.(\ref{lambda}%
), or in reduced units, $2\lambda \sum\limits_{n=0}^{N}\frac{1}{(2n+1)}=1.$
In the weak coupling limit of $\lambda \ll 1$, we have $N\gg 1$.

In general case of the arbitrary value of the parameter $\ell $ ($\ell
\simeq L/\xi _{0}$) the Eqs.(\ref{q(x)1}) are the convenient $\,$starting
point for the numerical calculation of function $J_{c}(\ell )$. We consider
here two limiting cases of $\ell \gg 1$ and $\ell \ll 1.$

For a long microbridge with $\ell \gg 1$ we shall seek a solution of Eq.(\ref
{q(x)1}) in the form $q(x)=\alpha x$. Substituting this $q(x)$ in Eq.(\ref
{q(x)1}), we find that $\alpha =2+O(1/\ell ).$ Calculating $J_{c}$ (\ref{Ic1}%
) with $q(x)=2x$, we find that the order parameter and the critical current
are

\begin{equation}
\Delta \left( z\right) =\Delta _{0}\left( \cos \frac{\varphi }{2}+i\frac{2z}{%
L}\sin \frac{\varphi }{2}\right) ,L\gg \xi _{0},  \label{delta2}
\end{equation}

\begin{equation}
I_{c}(L)=\frac{14}{3\pi ^{2}}\zeta \left( 3\right) I_{0}\frac{\hbar v_{F}}{%
T_{c}L}, L\gg \xi _{0}.  \label{Ic2}
\end{equation}
Expressions (\ref{delta2}), (\ref{Ic2}) coincide with the solution of GL
equations (with effective boundary conditions for the order parameter $%
\Delta $ ) for the clean superconducting microbridge \cite{KOJ}. Thus, our
microscopic approach with the boundary conditions (\ref{bound}) for Green
functions (not for $\Delta )$ gives the results of the phenomenological
theory at $L\gg \xi _{0}.$

For a short microbridge with $\ell \ll 1,$ in zero approximation on $\ell $
we have that $q(x)=0$ ($\Delta \left( z\right) =\Delta _{0}\cos \frac{%
\varphi }{2})$, $J_{c}=1$. Or, in dimension units, $I_{c}(0)=I_{0},$ in
agreement with formula (\ref{IKO}). The corrections to the zero
approximation depend on the value of the product $\ell N$. For very small $%
\ell \ll T_{c}/\omega _{D}$ ({\it i.e}. $L\ll a_{D}\simeq v_{F}/\omega _{D})$%
, the product $\ell N$ becomes small , although the $N\gg 1$ . As a result,
when $q(x,\ell )$ and $J_{c}(\ell )$ are calculated in the region $L<a_{D},$
the cutoff in the sums over $n$ must be taken into account. Apparently, when
the cutoff frequency appears explicitly but not through the value of $T_{c},$%
the applicability of the BCS theory becomes questionable. More rigorous
consideration, based on the Eliashberg theory of superconductivity \cite{Eli}%
, is needed in this case. Nevertheless, by using the BCS model with cutoff
frequency we suppose qualitatively to take into account the retardation
effects of electron-phonon coupling in our problem. In the domain, defined
by the following inequalities : $\ell N\ll 1$, $N\gg 1$, $\ell \ll 1$ , the
functions $b(x)$ (\ref{b1}) and $K(x)\,$ (\ref{K1}) have the asymptotes:

\begin{equation}
b(x)=4\lambda \ell N\{x\ln (\ell N)+x(C+\ln 2)+\frac{1}{4}[\ln (\frac{1+2x}{%
1-2x})+2x\ln (1-4x^{2})]\},  \label{b2}
\end{equation}

\begin{equation}
K(\left| x\right| )=-2\lambda N[\ln (2\ell N\left| x\right| )-1].  \label{K2}
\end{equation}
Where $C\simeq 0.577$ is the Euler constant. As it follows from Eqs.(\ref{b2}%
), (\ref{K2}) in this case the integral term in the equation (\ref{q(x)1})
is small, and calculating the critical current in the first approximation on
the small parameter $\ell N$ we can put $q(x)=b(x).$ As a result we have

\begin{equation}
\Delta (z)=\Delta _{0}(T)(\cos \frac{\varphi }{2}+ib(z/L)\sin \frac{\varphi
}{2}),L\ll a_{D},  \label{delta3}
\end{equation}

with $b(x)$ defined by expression (\ref{b2}\ ),

\begin{equation}
I_{c}(L)=I_{0}(1-\frac{8}{\pi \lambda }\frac{T_{c}L}{v_{F}}) , L\ll
a_{D}.  \label{Ic3}
\end{equation}

In the region $\{\ell \ll 1$ and $\ell N\lesssim 1\}$ the integral term in
the equation (\ref{q(x)1}) is numerically small as compared with the
out-integral term $b(x)$. By using in equation (\ref{q(x)1}) the $q(x)=b(x)$
as the rough approximation, we calculate the function $J_{c}(\ell )$ shown
in Fig.2.

For the case $\ell \ll 1,$ and $\ell N\gg 1,$ we can put $N=\infty $ in the
equation for $q(x)$ and $J_{c}(\ell ).$ The corrections to the critical
current in this region of the length $L$ can be estimated as

\begin{equation}
I_{c}\approx I_{0}\left( 1-const\frac{L}{\xi _{0}}\ln \frac{\xi _{0}}{L}%
\right) ,\quad a_{D}\ll L\ll \xi _{0.}  \label{Ic4}
\end{equation}

The expressions (\ref{Ic2}), (\ref{Ic3}) and (\ref{Ic4}) describe the
dependence of the critical current on the contact length in the limiting
cases of short and long channel. With increasing length $L$ the critical
current decreases. For ultrasmall $L\lesssim a_{D}$ the value of $\delta
I_{c}/I_{0}\sim (1/\lambda )(L/\xi _{0})$ directly depends on the BCS
coupling constant $\lambda $, and consequently it is sensitive to the
effects of the strong electron-phonon coupling.

\section{Conclusion}

We have studied the size dependence of the Josephson critical current in
ballistic superconducting microbridges. Near the critical temperature $T_{c}$, 
the Eilenberger equations have been solved selfconsistently. The closed
integral equation for the order parameter $\Delta $ (\ref{eqdelta}) and the
formula for the critical current $I_{c}$ (\ref{Ic}) are derived. Equations 
(\ref{eqdelta}), (\ref{Ic}) are valid for the arbitrary microbridge length 
$L$
in the scale of the coherence length $\xi_{0}\sim v_{F}/T_{c}$. In strongly
inhomogeneous microcontact geometry they replace the differential
Ginzburg-Landau equations and can be solved numerically. In the limiting
cases $L\gg \xi _{0}$ and $L\ll \xi _{0}$ the analytical expressions for $%
\Delta $ inside the weak link and for the $I_{c}\left( L\right) $ are
obtained. In Figure 3 the dependence of $I_{c}\left( L\right) $ on $L$ is
shown schematically. For long microbridge, $L\gg \xi _{0}$, the critical
current $\sim 1/L$ is in the correspondence with the phenomenological
consideration. The main interest presents the region $L\lesssim \xi _{0}$,
where the microscopic theory is needed. We have calculated the corrections
to the KO theory (\cite{KO}), which are connected with the finite value of
the contact size. The expression (2) for the Josephson current was obtained
in Ref.(\cite{KO}) in zero approximation on the contact size. For the $L\ll
\xi _{0}$ we obtained that $\delta I_{c}/I_{0}\sim -\frac{L}{\xi _{0}}\ln
\frac{\xi _{0}}{L}$ , where $I_{0}$ is the value of the critical current in
KO theory. Thus, the corrections to the value $I_{0}$ are small for $L\ll
\xi _{0}$, but the derivative $dI_{c}/dL$ has the singularity at $L=0$. This
singularity is smeared, if we take into account the finite value of the
ratio $T_{c}/\omega _{D}$. For ultra short microchannel, $L\lesssim
a_{D}\sim v_{F}/\omega _{D}$ (the dashed region in the Fig.3), the length
dependence of the critical current becomes $\delta I_{c}/I_{0}\sim -\frac{L}{%
\lambda \xi _{0}}$ ($\lambda $ is the constant of electron-phonon coupling).
In the very small microcontacts we have the unique situation, when the
disturbance of the superconducting order parameter can be localized on the
length $a_{D}$, making essential the effects of retardation of
electron-phonon interaction. The ballistic flight of electrons through the
channel is dynamical process with the characteristic frequency $\omega
_{0}\sim v_{F}/L$. For $L$ smaller then $a_{D}$ this frequency becomes
comparable with the Debye frequency $\omega _{D}$.

Thus, the critical current $I_{c}$ for the finite contact's size is smaller
then $I_{0}$. At the same time the normal state resistance $R_{N}$ of the
ballistic microchannel does not depend of the length $L$ and remains equal
to the Sharvin resistance $R_{0}$ (\ref{RSh}). As the result, the value of the
product $I_{c}R_{N}$ is not equal $\pi \Delta _{0}^{2}/4eT_{c}$ and depends
on the contact size. We have considered here the case of the quasiclassical
situation, $L\gg \hbar /p_{F}$. In quantum regime, $L\sim \hbar /p_{F}$, the
Sharvin resistance $R_{0}$ in formula (2) is substituted by quantized
resistance of the contact, as was firstly shown by Beenakker and Houten \cite
{Bee}. It follows from our consideration that for such small microcontacts
with $L\lesssim a_{D}$ the rigorous calculation of the Josephson current
requires to taking into account the retardation effects.

\section{Acknowledgments}

Authors are grateful to M. R. H. Khajehpour for useful discussions. This
work has been partly supported by the Institute for Advanced Studies in
Basic Sciences at Zanjan, IRAN.

\newpage

{\large Figure Captions}

Figure 1. Model of $S-c-S$ contact as narrow superconducting channel in
contact with bulk superconductors $S_{1}$ and $S_{2}.\ \ \ \ \ $

\bigskip Figure 2. Dependence of the critical current $I_{c}$ on the contact
length $L$ for the microbridge (solid line). The coupling constant $\lambda
=0.2$. For the comparison, the dependence $I_{c}\left( L\right) $ for S-N-S
contact ( $\lambda =0$ inside the channel) is shown (dashed line).

Figure 3. Dependence of the critical current on the length of the bridge.
The asymptotic behavior for short and long bridges are shown. The dashed
region corresponds to the ultrashort microbridge, $L\lesssim v_{F}/\omega
_{D}$.


\newpage

\begin{thebibliography}{99}
\bibitem{Jo}  B.D.Josephson, Weakly coupled superconductors,
Superconductivity, v.1, ed.R.D.Parks, Marcel Dekker, (1969).

\bibitem{AL}  L. G. Aslamazov and A. I. Larkin, {\it JETP Lett.}{\bf 9}  87
(1969). See also K. K. Likharev, {\it Rev. Mod. Phys.}{\bf 51}  101,(1979).

\bibitem{Poza}  M.Poza et al., Phys.Rev.B , {\bf 58}, 11173 (1998).

\bibitem{Mul}  C.J.Muller, J.M. van Ruitenbeek and L.J. de Jongh, Physica C 
{\bf 191,} 485 (1992) 

\bibitem{Na}  A.I.Yanson, G.Rubio Bollinger, H.E. van der Brom, N.Agrait,
J.M.van Ruitenbeek, Nature, {\bf 395,} 783 (1998).

\bibitem{Na2}  Hideaki Ohishi, Yukihito Kondo, Kunio Takayanagi, Nature, 
{\bf 395,} 780 (1998).

\bibitem{Ya}  I.K.Yanson, Fiz.Nizk.Temp., {\bf 1}, 141 (1978).

\bibitem{KO}  I. O. Kulik, A. N. Omelyanchouk, {\it Sov. J. Low Temp. Phys.} 
{\bf 4,} 142 (1978).

\bibitem{Sha}  Yu.V.Sharvin, Sov. Phys. JETP{\bf \ 21,}  655 (1965)

\bibitem{Ei}  G. Eilenberger, {\it Z. Phys.} {\bf 214,} 195 (1968)

\bibitem{La}  A.I.Larkin, Yu.N.Ovchinnikov, Sov.Phys.JETP {\bf 46},155
(1977).

\bibitem{KOJ}  I. O. Kulik, A. N. Omelyanchouk, Sov. Phys. JETP {\bf 41, }%
1071 (1976).

\bibitem{Eli}  G. M. Eliashberg, JETP, {\bf 38,} 969 (1960); {\bf 39}, 1437
(1960).

\bibitem{Bee}  C.W.J.Beenakker, H.van Houten. Phys.Rev.Lett, {\bf 66},
3056(1991).
\end{thebibliography}
\end{document}